\documentstyle[sprocl]{article}
\def\be{\begin{eqnarray}}
\def\ee{\end{eqnarray}}

\newcommand {\bnu} {{\vec\nu }}

\begin{document}
\title{GENERATION OF PULSAR GLITCHES: A SUPERFLUID CORE MODEL
\footnote{The Proceedings of the 18th Texas Symposium
on Relativistic Astrophysics;
eds. A.~Olinto, J.~Frieman, and D.~Schramm; World
Scientific Press. }}
\author{ A. SEDRAKIAN \&  J. CORDES}
\address{Center for Radiophysics and Space Research, Cornell University,\\ Ithaca, NY 14853}
\maketitle

\abstracts{
We show that the neutron star's crust-core interface acts as a potential 
barrier on the peripheral neutron vortices approaching this interface
and thus prevents their continuous decay on it in the course
of equilibrium state deceleration. 
The barrier arises due to the interaction 
of vortex magnetic flux with the Meissner currents set up 
by the crustal magnetic field at the interface. 
When the non-balanced part of the Magnus 
force reaches the value at which the vortices are able to annihilate 
at the interface, the rapid transfer of angular momentum from the 
superfluid spins-up the observable crust on short dynamical coupling times. 
}

Models of generation of pulsar glitches are supposed to
explain the following observational 
facts: (i)~the short spin-up time scales,
which are less than 120~s in the Vela pulsar 0833--45
and less than an hour in the Crab pulsar 0531+21; 
(ii)~the magnitudes of the jumps in the rotation and spin
down rates, $\Delta\nu/\nu \sim 10^{-8}-10^{-6}$ and 
$\Delta\dot\nu/\dot\nu  \sim 10^{-3}-10^{-2}$, respectively; and 
(iii)~the origin of the instability driving a glitch along with
characteristic intervals between glitches (typically of the order of 
several months to  years).
A number of existing generic models invoke as trigger crust- 
and core-quakes \cite{RUDERMAN} (discontinuous adjustments of the solid
crust to the gradually changing oblateness of the star as it spins down);
spontaneous quantum transitions of 
rotating superfluid between quasistationary eigenstates corresponding to
different eigenvalues of total angular momentum \cite{PACKARD}; 
collective unpinning of a large number ($\sim 10^{13}$) of vortices 
in the neutron star crusts \cite{ANDERSON,ALPARETAL} and 
thermal instabilities \cite{LINK}. The increasing bulk of observational
evidence provides good chances of discrimination between the theoretical
models in the future.  

Here we shall give a brief account of a new trigger mechanism 
for generation of pulsar glitches in neutron star's superfluid core. 
The complete discussion will be given in ref. 6.
In the interjump epoch a neutron star is  
decelerating; consequently the vortex lattice in the superfluid core is 
expanding and the peripheral vortices attempt to cross the crust-core 
boundary. The crust-superfluid core interface acts as a 
potential barrier on the vortices in the superfluid core that approach this 
boundary. The barrier arises due to the magnetic interaction between the 
crustal magnetic field $H_0$ (which penetrates the 
superconducting core exponentially  
within a scale $\delta_p$ - the penetration depth) and quantum vortices
whose magnetic field  is governed by the generalized London equation
\be\label{LONDON} 
\delta_p^{-2}\,\vec\nabla \times (\vec\nabla \times \vec B_v)+\vec B_v = 
\bnu_p\, \, \Phi_0\delta^{(2)}(\vec r-\vec r_p)
+\bnu_n\,\Phi_1\delta^{(2)}(\vec r-\vec r_n).
\ee
Here $\bnu$'s are circulation unit vectors; subscripts $p$ and $n$ 
refer to protons and neutrons; $\Phi_0$ is the flux quantum carried 
by proton vortices; and $\Phi_1$ is the non-quantized flux of neutron 
vortices due to the entrainment effect (i.e. the effect
of superconducting proton mass transport by the neutron superfluid 
circulation, see ref. 7). The total field $\vec B$ is the 
superposition of $\vec B_v$ of eq. (\ref{LONDON}) and crustal filed
$\vec B_{cr} = \vec H_0 {\rm exp} ( -\vec r\cdot\vec n/\delta_p)$, 
($\vec n$ being the normal of the interface).

Suppose that  the interface is the $(yz)$-plane of a Cartesian system of 
coordinates  and the vectors of vortex circulation are in the positive 
$z$-direction. The half-plane $x<0$ corresponds to the crust 
while $x>0$ corresponds to the superfluid core. The knowledge of magnetic 
field distribution with the boundary condition at the interface $B_z=H_0$,
allows one to calculate the relevant part of the Gibbs free
energy of the system $G = F - (4\pi)^{-1}\int \vec B\cdot \vec H_0 dV$, 
where  the free-energy is  
\be\label{GIBBS} 
F= 
\frac{\delta_p^2}{8\pi}\int\left[\,\vec B\times (\vec \nabla \times \vec B) \,\right]\cdot d\vec S +\frac{\delta_p^2}{8\pi}
\int\!\vec B\cdot[~\delta_p^{-2}\vec B+\vec\nabla\times(\vec \nabla 
\times \vec B)~]~dV .
\ee
The force per single vortex of effective flux $\Phi_*(\equiv \Phi_0; 
\, \Phi_1)$  derived from  $G$  is 
\be\label{FORCE} 
f(x ) 
= \frac{\Phi_*}{4\pi}\left[\frac{H_0}{\delta_p}\,
e^{-x/\delta_p
}-\frac{\Phi_*}{2\pi\delta_p^3}\,
 K_1\left(\frac{2\, x}{\delta_p}\right)\right].
\ee
Here $K_1$ is the modified Bessel function.
The first term in equation (\ref{FORCE}) is the 
repulsive force acting between the vortex magnetic flux  and the crustal 
magnetic field. It can also be interpreted as a Lorentz force resulting 
from superposition of velocity fields of the vortex and the surface 
Meissner currents. The second term is the attractive force acting 
between the vortex and the interface. 
For large distances the repulsive term dominates;
(the attractive one goes to zero faster
since $K_1(2x)\propto \sqrt{\pi/4x} ~e^{-2x}$ for $x\to \infty$).
For small distances the second (attractive) term in 
eq. (\ref{FORCE}) dominates.
Thus the repulsive part of the vortex - crust-core interface 
interaction, which dominates at large distances, acts as a potential 
barrier  on a vortex approaching the boundary. 
The Magnus force is then balanced by the vortex-interface force.
When the disbalance between 
this forces reaches the value at which the vortices annihilate 
at the interface, the resultant rapid transfer of angular 
momentum from the superfluid spins-up the crust.

Let us next estimate the interaction. 
For relevant densities, $\rho\simeq 2\times 10^{14}$ g cm$^{-3}$, 
we have $\delta_p \simeq 100$ fm (e. g. ref. 7)
and  assuming a conventional value for the crustal magnetic field 
$H_0 = 10^{12}$ G, we find a maximal repulsive force   
$f^{\rm max} =3.13\times 10^{12}$ dyn cm$^{-1}$. 
In general, the magnitude of the crustal magnetic 
field at the  crust-core interface for different objects 
can vary in a reasonable range $10^9< H_0<10^{13}$ G.
We find that, the maximal force increases (decreases) by two 
orders of magnitude when the crustal magnetic field is increased 
(decreased) by an order of magnitude.

Further progress needs to specify the ground state structure 
of vortices in the superfluid core. For the vortex cluster 
model of ref. 7, 8 (no residual field when the superconducting 
state sets on or a complete Meissner expulsion of residual field)
the number of proton vortices per neutron vortex  
interacting with the interface is $N\le 283$ and 
the maximal force is $f^{\rm max}_C = N~ f^{\rm max} 
= 8.9\times 10^{14}$ dyn cm$^{-1}$. It should sustain 
the Magnus force excess (i.e. the part which is non-balanced  
 along  the normal $\vec n$) acting on a neutron vortex 
$\delta f^M =3.23\times 10^{17}~(\delta\omega_s/{\rm s}^{-1}) 
~ {\rm dyn}~{\rm cm}^{-1}, $
where $\delta\omega_s$ is the angular velocity difference between the 
superfluid and the normal components.\footnote{The neutron star model 
used in our estimates is discussed in detail in ref. 8.} 
From the  balance condition $f^{\rm max}_C = \delta f^M$,  the maximal 
departure that can be sustained  by the boundary force on the 
cluster is  $\delta\omega_s^{\rm max} \simeq 0.003$ s$^{-1}$.
Then the angular momentum conservation
in a Vela-type glitch implies that the ratio of the moment 
of inertia of the superfluid region to the normal component 
should be $I_s/I_n = 0.023$.
This value is close to a previous estimate  of the ratio $I_s/I_n = 0.02$  
for a superfluid shell at the crust-core boundary with short dynamical 
coupling times $(\le 120$ s) \cite{PAPER2}.  

\section*{Acknowledgments}
One of us (A.S.) gratefully acknowledges a research grant from the Max-Kade-Foundation, NY.

\end{document}